\documentclass[fleqn,usenatbib]{mnras}

\usepackage{newtxtext,newtxmath}

\usepackage[T1]{fontenc}

\DeclareRobustCommand{\VAN}[3]{#2}
\let\VANthebibliography\thebibliography
\def\thebibliography{\DeclareRobustCommand{\VAN}[3]{##3}\VANthebibliography}

\usepackage{graphicx}	
\usepackage{amsmath}	
\usepackage{amssymb}	

\title[Analysis of contamination in LAMOST-MRS spectra]{Analysis of the possible satellite contamination in LAMOST-MRS spectra}

\author[M. Kovalev et al.]{
Mikhail Kovalev$^{1,2,3}$\thanks{E-mail: mikhail.kovalev@ynao.ac.cn},
Olivier R. Hainaut$^{4}$,
Xuefei Chen$^{1,2,5}$,
Zhanwen Han$^{1,2,5}$
\\
$^{1}$Yunnan Observatories, China Academy of Sciences, Kunming 650216, China\\
$^{2}$Key Laboratory for the Structure and Evolution of Celestial Objects, Chinese Academy of Sciences, Kunming 650011, China\\
$^{3}$Sternberg Astronomical Institute of the M. V. Lomonosov Moscow State University, Leninskie Gory, Moscow 119992, Russia\\
$^{4}$European Southern Observatory, Karl-Schwarzschild-Stra\ss e 2, 85748 Garching bei München, Germany\\
$^{5}$Center for Astronomical Mega-Science, Chinese Academy of Sciences, 20A Datun Road, Chaoyang District, Beijing 100012, China\\
}

\date{Accepted ----. Received ----; in original form ----}

\pubyear{2023}

\def\kms{\,{\rm km}\,{\rm s}^{-1}}

\newcommand{\teff}{{T_{\rm eff}}}
\newcommand{\rv}{{\rm RV}}

\def\vsini{V \sin{i}}
\def\logg{\log{\rm (g)}}
\def\snr{\hbox{S/N}}
\def\drv{\hbox{|$\Delta$ RV|}}
\def\imp{f_{\rm imp}}

\begin{document}
\label{firstpage}
\pagerange{\pageref{firstpage}--\pageref{lastpage}}
\maketitle

\begin{abstract}
We present the detection of false positive double-lined spectroscopic binaries candidates (SB2) using medium-resolution survey (MRS) spectra from the one time-domain field of LAMOST data release 10 (DR10). The secondary component in all these binaries has near zero radial velocity and solar-like spectral lines. Highly likely this is light from the semi-transparent clouds illuminated by the full Moon. However we also suspect that partially this contamination can be caused by a solar light reflected from the surface of low-orbital artificial satellites launched in the beginning of 2022. We found several possible contaminant candidates using archival orbital data. We propose measures to reduce risk of such contamination for the future observations and methods to find it in archived ones.     
\end{abstract}

\begin{keywords}
binaries : spectroscopic -- techniques : spectroscopic
\end{keywords}



\section{Introduction}


Since the launch of the Sputnik-1 in 1957, we can see artificial satellites flying in the night sky. Such observations can be very useful for Earth-related science (i.e. determination of the geopotential), although for astrophysics, satellites can be an obstacle.
This problem has become more serious with the start of the active populating of low earth orbits, which now host many thousands of telecommunication satellites, which form huge constellations. In most pessimistic scenario such intensive commercialisation of space can be the end of all ground base astronomy.
\par
In spectroscopic observations, the flyby of an artificial satellite will result as a fake spectroscopic binary, where contamination will be visible as a solar-like spectral component. For low orbit satellites, the line of sight velocity ($\rv$) is near zero when the satellite rise close to culmination, but the transverse velocity is very high, so contamination lasts much less than a second for typical values of field of view. Thus for a typical bright astrophysical target, contamination is usually negligible and only relatively faint objects are affected \citep{bassa2022}.
\par
\cite{bincat} identified many double-lined spectroscopic binary (SB2) candidates in LAMOST (Large Sky Area Multi-Object fiber Spectroscopic Telescope) MRS \citep{lamostmrs}. However some of them can be false positives, which can be identified by taking advantage of multiple observations in a time domain sub-survey. Here we present results for one particular field, where these false-positive SB2s can be caused by satellite contamination.
\par
The paper is organised as follows: in Sections~\ref{sec:obs} and \ref{sec:methods}, we describe the observations and methods. Section~\ref{results} presents our results. In Section~\ref{discus} we discuss the results. In Section~\ref{concl} we summarise the paper and draw conclusions.

\section{Observations}
\label{sec:obs}

LAMOST is a 4-meter quasi-meridian reflective Schmidt telescope with 4000 fibers installed on its $5\degr$ FoV focal plane. These configurations allow it to observe spectra for at most 4000 celestial objects simultaneously (\cite{2012RAA....12.1197C, 2012RAA....12..723Z}).
 For the analysis in this paper, we downloaded all available time-domain DR10 spectra from \url{www.lamost.org/dr10/v0/} observed within the field ``{\rm TD164021N701415T01}".	We use the spectra taken at a resolving power of $R=\lambda/ \Delta \lambda \sim 7\,500$. Each spectrum is divided on two arms: blue from 4950\,\AA~to 5350\,\AA~and red from 6300\,\AA~to 6800\,\AA.~During the reduction, heliocentric radial velocity corrections in range of $\rv_h=-5,-2~\kms$ were applied to all spectra. We convert the wavelength scale in the observed spectra from vacuum to air using \texttt{PyAstronomy} \citep{pya}. Observations are carried out in MJD=59676.8-59692.8 days, spanning an interval of 16 days. 
 We selected only spectra stacked for whole night\footnote{ Each epoch contains seven short 20 min individual exposures, which were stacked to increase $\snr$} and apply a cut on the signal-to-noise ($\snr>=20$). In total we have $5625$ spectra from $1323$ targets. The number of epochs varies from 2 to 4 per target, as very noisy epochs were not selected for some targets.

\section{Methods} 
\label{sec:methods} 
\label{sec:specfit}

We use the same spectroscopic models and method as \cite{tyc,bincat} to analyse individual LAMOST-MRS spectra, see very brief description below.
The normalised binary model spectrum is generated as a sum of the two Doppler-shifted normalised single-star spectral models ${f}_{\lambda,i}$\footnote{they are designed as a good representation of the LAMOST-MRS spectra}, scaled according to the difference in luminosity, which is a function of the $\teff$ and stellar size. We assume both components to be spherical and use the following equation:    

\begin{align}
    {f}_{\lambda,{\rm binary}}=\frac{{f}_{\lambda,2} + k_\lambda {f}_{\lambda,1}}{1+k_\lambda},~
    k_\lambda= \frac{B_\lambda(\teff{_{,1}})~R^2_1}{B_\lambda(\teff{_{,2}})~R^2_2}
	\label{eq:bolzmann}
\end{align}
 where $k_\lambda$ is the luminosity ratio per wavelength unit, $B_\lambda$ is the black-body radiation (Plank function), $\teff$ is the effective temperature and $R$ is the stellar radius. Throughout the paper we always assume the primary star to be brighter one. In comparison with \cite{bincat} we directly use the ratio of stellar radii $q$ as a fitting parameter, instead of the mass ratio with difference of the surface gravity $\logg$.   
\par
\par 
Each spectrum is analysed with the single and binary spectral model, thus we can calculate the difference in reduced $\chi^2$ between two solutions and the improvement factor ($\imp$), computed using Equation~\ref{eqn:f_imp} similar to \cite{bardy2018}. This improvement factor estimates the absolute value difference between two fits and weights it by the difference between the two solutions.

\begin{align}
\label{eqn:f_imp}
f_{{\rm imp}}=\frac{\sum\left[ \left(\left|{f}_{\lambda,{\rm single}}-{f}_{\lambda}\right|-\left|{f}_{\lambda,{\rm binary}}-{f}_{\lambda}\right|\right)/{\sigma}_{\lambda}\right] }{\sum\left[ \left|{f}_{\lambda,{\rm single}}-{f}_{\lambda,{\rm binary}}\right|/{\sigma}_{\lambda}\right] },
\end{align}
where ${f}_{\lambda}$ and ${\sigma}_{\lambda}$ are the observed flux and corresponding uncertainty, ${f}_{\lambda,{\rm single}}$ and ${f}_{\lambda,{\rm binary}}$ are the best-fit single-star and binary model spectra, respectively, and the sum is over all wavelength pixels.

\section{Results}
\label{results}


We carefully checked the quality of the spectral fits through visual inspection of the plots. Several spectra were selected as SB2 candidates using criteria formulated in \cite{bincat}, although this selection was not complete as these criteria prioritise purity. This study is focused on possible satellite contamination, so we introduce a new selection of the fitted parameters, like $\rv$ and improvement factor, see Table~\ref{tab:cuts}. 
\begin{table}
    \centering
    \caption{Selection criteria}
    \begin{tabular}{l}
\hline
Selection criteria\\
\hline
|$\rv_2$+5|<15 $\kms$\\
$\vsini_2$<30 $\kms$\\
$f_{\rm imp}>0.1$\\
|$\teff_2-5777$|<700 K\\
\hline

    \end{tabular}
    \label{tab:cuts}
\end{table}

\begin{figure}
    \centering
    \includegraphics[width=\columnwidth]{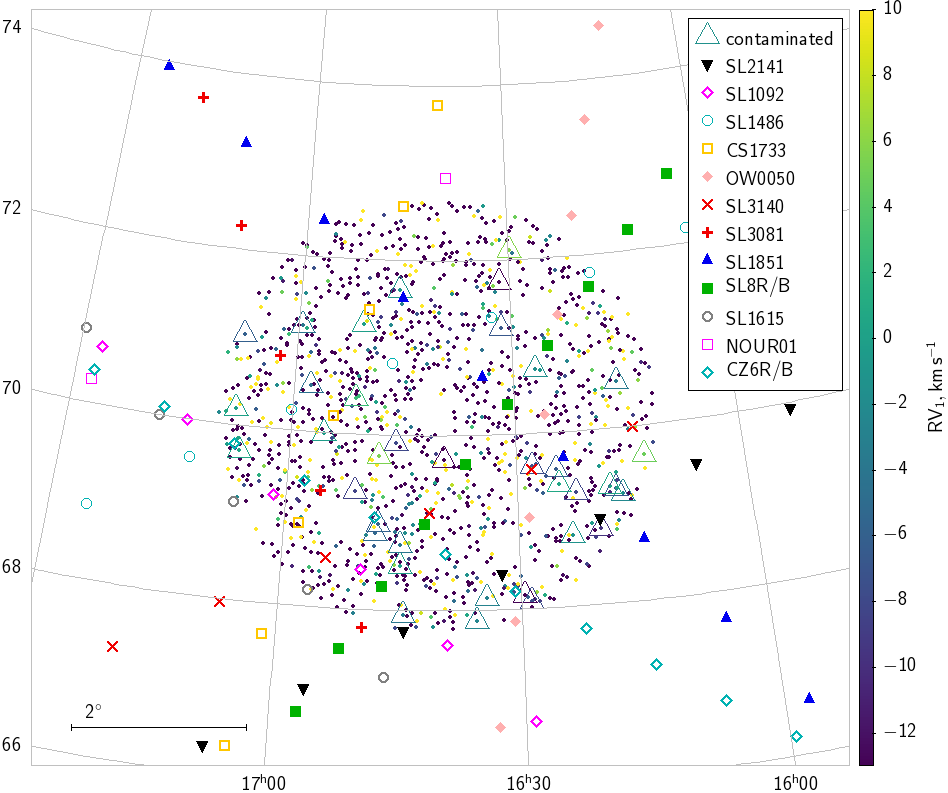}
    \caption{Contaminated field observed at night MJD=59685.8 d. Selected contaminated spectra are highlighted by open triangles. Datapoints are color-coded by the $\rv_2$. Computed satellite positions are shown for two seconds intervals. SL = Starlink, CS = Cosmos, OW = OneWeb.}
    \label{fig:field}
\end{figure}

Out of four epochs, one with MJD=59685.8 d has significantly more selected candidates, so we explored it more carefully. Thus we keep only stars that appear as a regular single star in all epoch except MJD=59685.8 d. In total we left with 37 SB2 candidates, with a secondary component at $\rv_2\sim 0~\kms$.  They are marked as open triangles on Figure~\ref{fig:field}.

\begin{figure*}
	\includegraphics[width=0.9\textwidth]{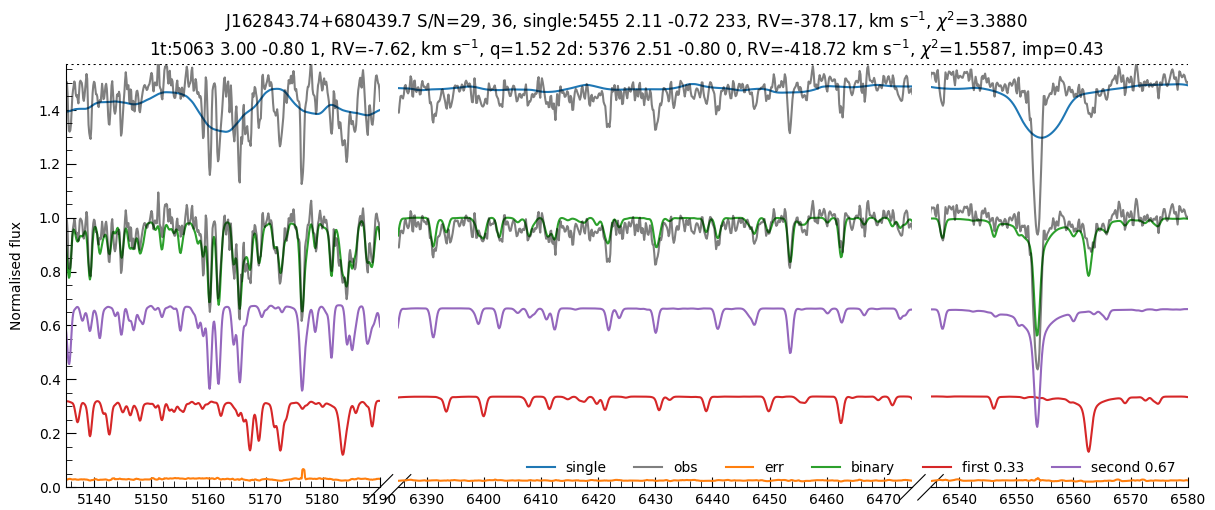}
	\includegraphics[width=0.9\textwidth]{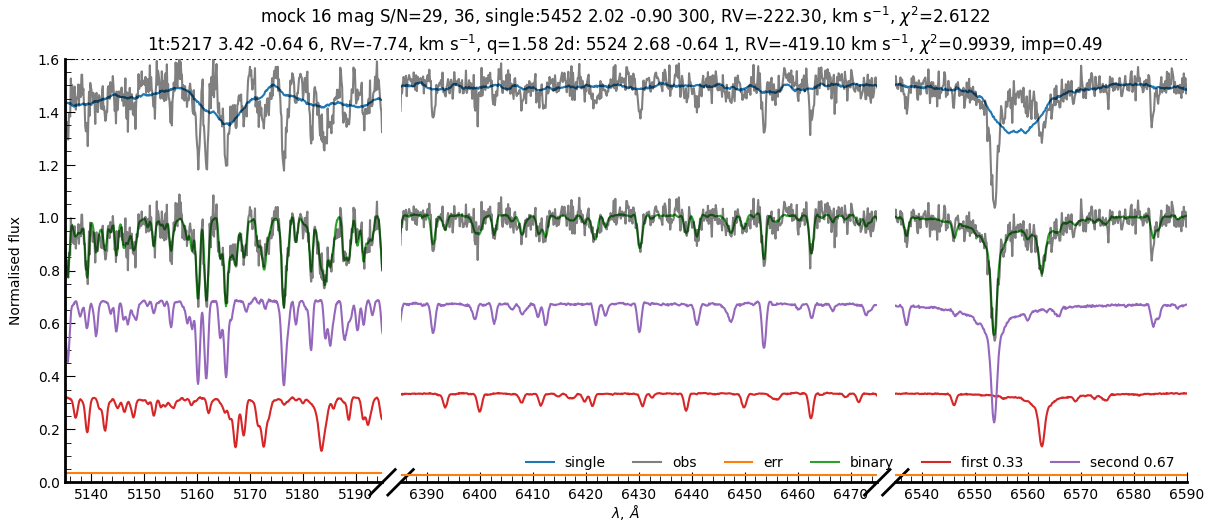}
	\includegraphics[width=0.9\textwidth]{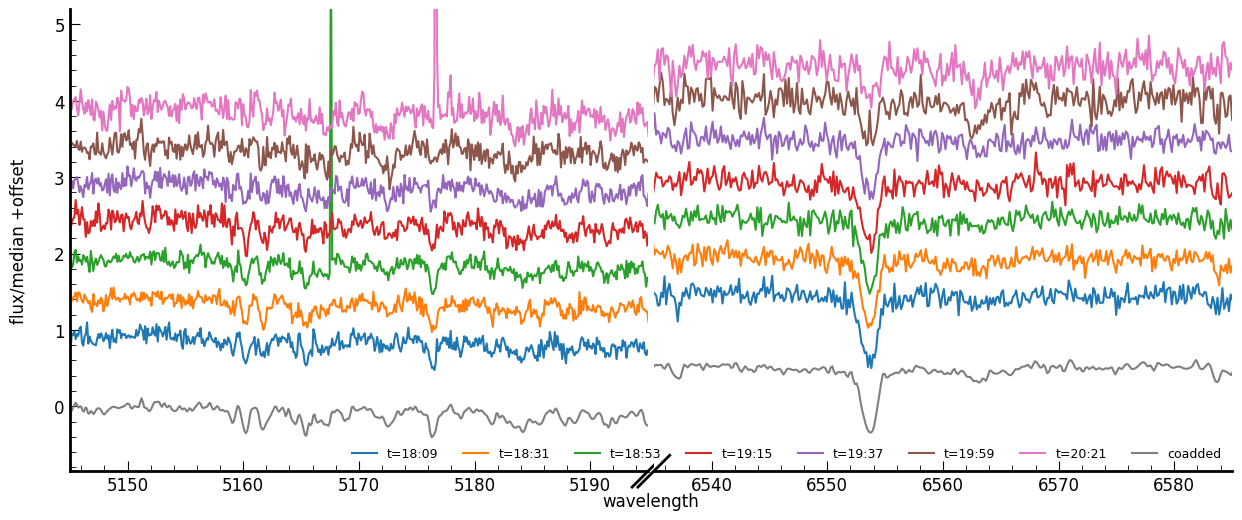}
    \caption{Example of the co-added spectrum of J162843.74+680439.7 that was fitted by the single-star and binary models on the top panel and fitting of the mock spectrum contaminated by satellite with $V_{\rm eff}=16$ mag in the middle panel. The bottom panels shows individual 20 minute exposures, where times indicated in the legend show the median UTC during the exposure.}
    \label{fig:spfit}
\end{figure*}

We show the most clear example J162843.74+680439.7 ($G=14.55$ mag) with very large $\drv\sim410~\kms$ in Fig.~\ref{fig:spfit}. In the top panel we show fits of the co-added spectrum by the single-star and binary model. The single-star model obviously failed to fit double-lined spectrum, while binary models fit the primary component (67 per cent) at $\rv_1=-418.72~\kms$ and catch another additional spectrum component (33 per cent) at $\rv_2=-7.62~\kms$. In the middle panel we show fitting results for a mock spectrum of J162843.74+680439.7 contaminated by solar spectrum of $V=16$ mag, where we applied Gaussian noise according to $\snr$. As you can see both panels are very similar. 
 In the bottom panel we show all seven short 20 min exposures spectra before coaddition. It is clear that contamination happened at UTC times $t=$19:59 and $t=$20:21 as these two exposures have an additional spectral component, which has a brightness comparable with the main target. When all exposures were co-added we got the double-lined spectrum with significantly smaller noise. 

\par
In the other candidates contamination is not that clearly visible as they have smaller $\drv$. The majority of the candidates have $G\sim14.5$ and $\snr_{\rm red}<50$ in the co-added spectrum, thus probably for brighter targets contamination was negligible and comparable to the noise level. 

\section{Possible source of contamination}
\label{discus}

\begin{figure}
	\includegraphics[width=8.8cm]{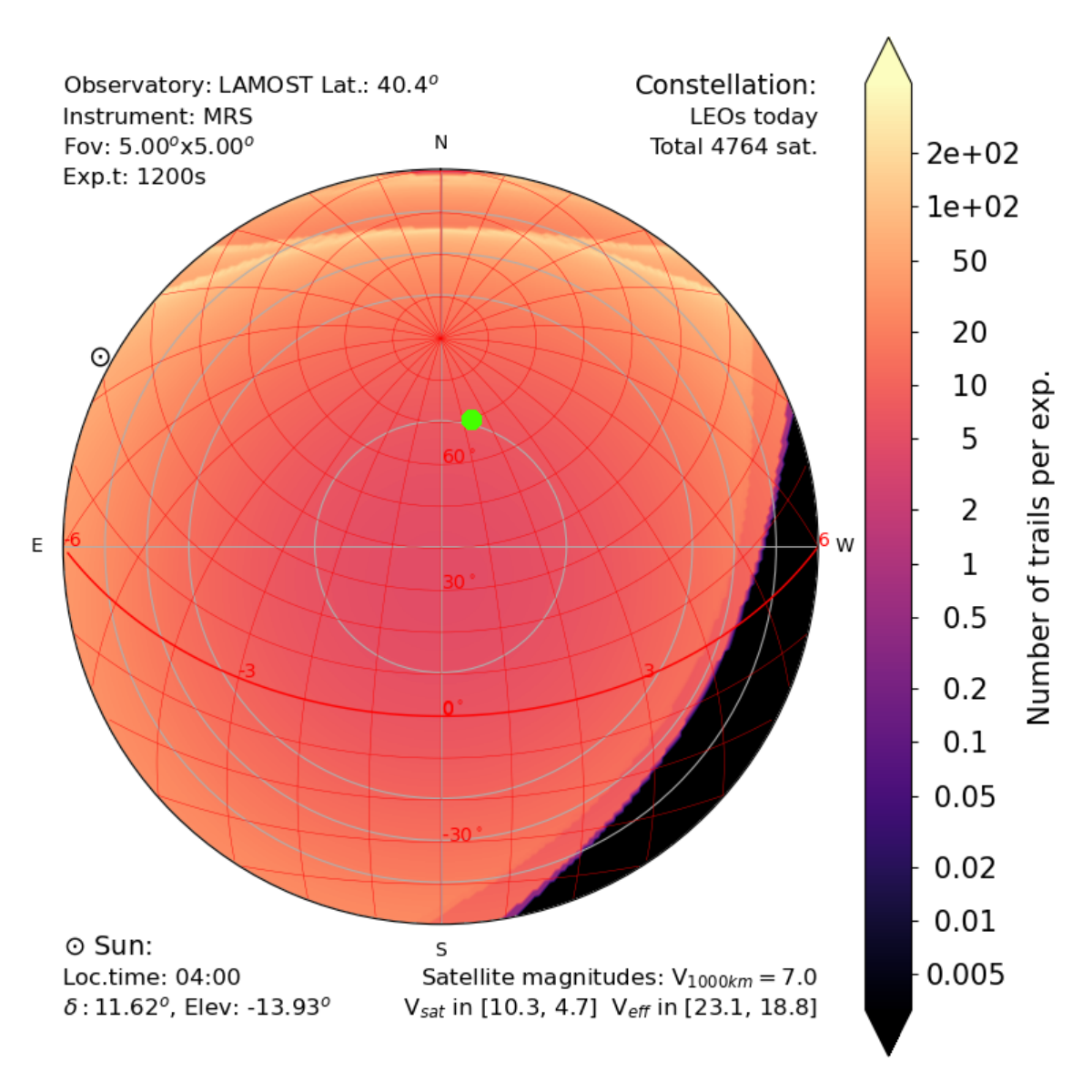}
    \caption{Map of the sky above the observatory, showing the telescope pointing (green dot) during the contaminated twilight observation. The color scale shows the number of satellites crossing a LAMOST field-of-view during a 1200s exposure. The considered satellites include pre-constellation low-Earth orbit satellites and the Starlink and OneWeb satellites in orbit at the time of the observations. The black area mark the zone of the sky where the satellites are in the shadow of the Earth. Their modelled magnitudes are in the 4.7--10.3 range. Accounting for the trailing, their effective magnitude for LAMOST is in the 18.8--23.1 range, i.e well fainter than the limiting magnitude of the instrument. }
    \label{fig:bag}
\end{figure}
\begin{figure}
	\includegraphics[width=8.8cm]{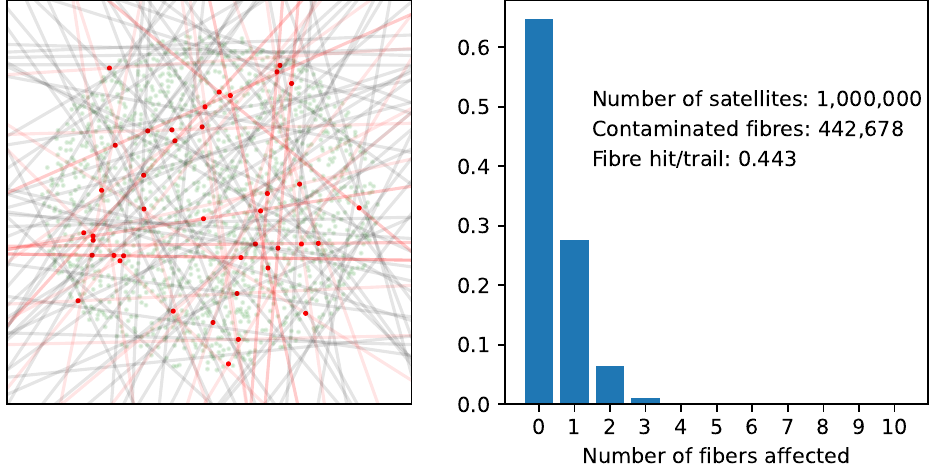}
    \caption{Left: Examples of 100 satellite trails shot randomly through a LAMOST field of view (right panel). Fibres free of contamination are marked in green, while those affected by a satellite are in red. The trails hitting at least one fibre are in red. Right: Histogram of the fraction of satellite trails as a function of the number of fibres hit, from 1 million random trails as in previous figure. On average, a trail will hit 0.44 fibre. }
    \label{fig:fov}
\end{figure}

Highly-likely such contamination was caused by the clouds, illuminated by the full Moon, which significantly increased sky background in the spectra. Unfortunately, sky subtraction failed to completely remove it during the spectral reduction,  see the last two individual 20-minute exposures in the bottom panel of Fig.~\ref{fig:spfit}. This can explain such solar-like spectral component very well as sky becomes brighter as sun is rising. At the moment of the end of observation it's height was around $-12\degr$. This also supported by the fact that contamination is visible only in relatively faint targets. Nevertheless we decided to test other possible sources of contamination.
\par
 We checked if this contamination can be due to a solar system object. We used Minor Planet Center checker\footnote{\url{https://minorplanetcenter.net/cgi-bin/mpcheck.cgi}} to check 1284083 known objects and found none of them brighter than 18 mag in our field. With slightly larger search radii we found comet C/2019 K7 (Smith) with coordinates $\alpha$=16:20:45.9,~$\delta=+67^\circ 56' 19"$, although it is unlikely to be our contaminant, because otherwise it will be visible in all exposures, as it moves very slowly.


In order to investigate whether this contamination could have been caused by a satellite passing through the field of view, we verified that, at the time of the observations, low-Earth orbit (LEO) satellites were illuminated by the Sun. This was tested using the formalism described in \citet{bassa2022} for generic LEOs as well as for Starlink and OneWeb satellites.

To evaluate the number of fibres typically affected by a satellite trail, a million trails, randomly positioned, were shot through a realistic LAMOST field of view. For the considered field, 1324 fibres had object of interest (with suitable S/N$>20$) over the 4000 fibres of the instrument, so 1324 fibres were considered in this experiment. 
A trail is considered to affect a fibre if the impact distance is less than 3$''$, which accounts for the radius of the fibre (whose diametre is 3.3$''$) and the width of the trail, which is set to 2$''$ accounting for the seeing and the marginally resolved satellite. For each trail, the number of fibres affected was counted. Figure~\ref{fig:fov} illustrates this for 100 trails on the left panel, and displays a histogram of the number of fibres affected on the left panel. This method is the same that was used to evaluate (Michevat priv.comm.) the impact on 4MOST, a similar spectrograph built at ESO \citep{4most2019}. About 64\% of the trails hit no fibre and while 0.01\% of the satellites hit 7 fibres, a trail will hit 0.44 fibres on average. As 37 fibres were contaminated, this suggests up to $\sim$80 satellites crossed the $5\degr$ field of view during the exposure. These numbers should be taken with a fairly large uncertainty, as the seeing and the width of the trail will cause the number of fibres affected to be larger, but the contamination for a larger impact distance will be smaller.

To estimate the visual magnitude of the satellite causing the contamination, one must estimate the level of contamination of the spectra, and take into account the effect of motion of the satellite. With typical angular velocities of the order of 1\degr~$s^{-1}$ at zenith, a LEO satellite spends only a few milliseconds $t_{\rm eff}$ crossing the fibre  during the total exposure time $t_{\rm exp} = 1200$ s. The apparent magnitude $m$ of the object can be estimated from its effective magnitude $m_{\rm eff}$ measured on the spectrum,
\begin{equation}
    m = m_{\rm eff} + 2.5 \log_{10} \frac{t_{\rm eff}}{t_{\rm exp}} 
      = m_{\rm eff} + 2.5 \log_{10} \frac{r_{\rm fibre}}{\omega_{\rm sat} t_{\rm exp}} , \label{eq:meff}
\end{equation}
where $r_{\rm fibre} = 3.3''$ is the angular diameter of a fibre on the sky. 
Using the method in \citet{bassa2022}, the angular velocity of the satellite in the direction of observations was estimated for Starlink ($0.66$\degr~$s^{-1}$) and OneWeb ($0.30$\degr~$s^{-1}$) satellites. 

The effective magnitude can be estimated from the contamination. The S/N of the $G \sim 14.5$ was up to 50 in the co-added spectrum, corresponding to $\sim 20$ in the individual 1200s exposures. To be noticeable, the contamination must have S/N $> 5$ (which corresponds to $G\sim16$), and to be detectable at all, S/N$>2$ ($G\sim 17)$.
Combining these pieces of information, Eq.~\ref{eq:meff} gives visual magnitudes $\sim$1--2.
Fainter satellites will not be detected.

As of the time of the observations, about 4500 satellites were present on LEOs (roughly 2000 pre-existing, and 2002 Starlink\footnote{
\label{JMcDSL}
Jonathan McDowell’s Starlink web page
\url{https://planet4589.org/space/con/star/stats.html}
} and 426 OneWeb\footnote{
\label{JMcDOW}
Jonathan McDowell’s OneWeb web page
\url{https://planet4589.org/space/con/ow/stats.html}
} from recently launched mega-constellations). 
Using the method of \citet{bassa2022}, this results in $\sim$ 15 satellite trails per exposure during long twilight, as illustrated in Fig.~\ref{fig:bag}. 
This number is much too low to explain the observed contamination. Furthermore, the magnitudes of the satellites differs widely (some of them, such as HST or ISS can be as bright as $V$ -5 to 2), but the bulk of the Starlink satellites are in the 5.6--7.2 range \citet{mallama2021a, mallama2021b} and OneWeb in the 7--9 range \citet{mallama2020}, ie well below the reach of the spectrograph. 
\par
We also checked Satellite Track Predictor (STP)\footnote{\url{http://www.astro.amu.edu.pl/STP}} for time interval UTC=19:30, 20:30 and found that 12 bright satellites with $V\leq6$ mag crossed our field. We show their tracks in Fig.~\ref{fig:field}. STP reports that errors can be up to $0.1-0.5\degr$ for sky-positions and $\sigma_V=2$ mag for brightness, so some of these satellites (like Starlink and Cosmos with reported $V=4$ mag) can be bright enough to cause contamination.  
\par
In the week after their launch, the satellites appear as a train, or like a string of pearls while they slowly disperse in elongation along their very low orbit. During that phase, they appear much brighter than when on their operational orbit, because of the shorter distance to the observer, and because the configuration and attitude of the satellites are different than when in operations. In the days of the earliest Starlink launches, they could be as bright as mag $\sim 0$. Since then, the operator has modified the attitude of the satellites so that they are much dimmer, in the 1--3 range most of the time\footnote{Although very bright (up to $V\sim 0$ mag) and short ($\sim1$ sec) flashes are possible. First author saw them several times.}. A batch of satellites launched with one rocket consist typically of 60 satellites. In order to test whether such a train of recently launched satellites could have crossed our field of view, Two Line Elements (TLEs), the orbital elements of the satellites, were retrieved for the date of the observations using CelesTrack \footnote{\url{https://celestrak.org/NORAD/archives/request.php}}. Using the {\sc skyfield}\footnote{\url{https://rhodesmill.org/skyfield/}} package, the visibility of the satellite was verified, from LAMOST for the time of the observations. It appears that a series of Starlink satellites from the 2022 Feb. 21 launch$^{\ref{JMcDSL}}$ crossed the sky during the exposure. While their tracks, as computed by us, are in the general vicinity of our observation, they does not cross the field of view. However, the TLEs are notoriously not very accurate --especially at a phase when the operator frequently adjust the orbit, and our method to compute the satellite position is not verified. At that time, the satellites were at an altitude of 350km, with a magnitude in the 1--2 range. The apparent angular velocity of these satellites was $\omega \sim 1.0$\degr~$s^{-1}$, which leads to effective magnitudes $m_{\rm eff} \sim 16$--17, i.e. in the range of the contamination.
Therefore, we suggest that the observations can be theoretically, "photobombed" by a train of Starlink satellites on their low, parking orbit, although contamination by clouds is more likely. 
\par
In the future, the number of satellites in mega-constellations is likely to grow significantly. Assuming 65\,000 satellites (as in \citet{bassa2022}), this would result in a typical 1200s exposure being crossed by about 200~satellite trails, potentially resulting in $\sim 260$ fibres contaminated per exposure taken during long twilight (3\% of the fibres). However, the limiting magnitude of the LAMOST-MRS instrument for 1200s exposure is $V\sim 15$ ($5\sigma$). Converting the apparent magnitudes of the satellites (using the crude photometric model described in \citet{bassa2022}) into effective magnitudes, these will be in the 18 to 23 range (depending on the satellite's orbit and altitude and azimuth), well below the limit of LAMOST-MRS, even accounting for a possible 1~mag error on the photometric model.
As usual, it is important to note that once the sun dips far enough under the horizon, most of the satellites fall in the shadow of the Earth. This problem is therefore only critical during the first and last hours of the night.

While the satellites on operational orbits will not be a major concern for LAMOST, the compact trains of very low satellites can affect the observations. The probability of such a train crossing a telescope field of view is low, but considering that constellations will need to be regularly replenished, new satellites will need to be continuously launched. Considering 100\,000 satellites with a life-time of 5~years, this would result in about one launch per day (each with 60 satellites). If the satellites stay one month in low orbit, this would result in about 60 trains in orbit, at various stage of dispersion. It is therefore important that the satellite operators also keep the brightness of the satellites to the absolute minimum possible during their stay on transit orbit. The changes of satellite attitude implemented by Starlink illustrate the improvements than can be made.

\section{Conclusions}
\label{concl}
 
We successfully detected false-positive SB2 candidates in the LAMOST-MRS spectra.
The secondary component in all these binaries have near zero radial velocity and solar-like spectral lines. Highly likely this is light from the semi-transparent clouds illuminated by the full Moon. However we also suspect that partially this contamination can be a solar light reflected from the surface of low-orbital artificial satellites launched in the beginning of 2022. We found several possible contaminant candidates using archival orbital data from CelesTrack and STP web service.
Unfortunately results presented in this paper cannot definitely confirm satellites as contaminant, as other sources like clouds and problem with sky subtraction will have similar effect on the spectral observations. 

To identify and remove such contamination we recommend analysis of all spectra taken during twilight, assuming a binary spectrum model, where one component has solar-like spectrum with radial velocity in the range $\drv=-10,+10~\kms$.
Also the short exposures should be carefully checked prior the co-addition to avoid the production of false double-lined spectra with contaminated exposures. 
\par 
During the scheduling of the observation one should consider possibility of the contamination by the bright "train" of newly launched satellites and avoid observations near the twilight if possible. Also we recommend to take additional image of the observed field, to reliably identify possible satellite tracks.

\section*{Acknowledgements}
MK is grateful to his parents, Yuri Kovalev and Yulia Kovaleva, for their full support in making this research possible. We thank Hans B{\"a}hr for his careful proof-reading of the manuscript. We thank Zhang Haotong and Luo A-Li for useful discussions. We thank Dr. Nikolay Emelyanov for providing the link to Minor Planet Center Checker. We thank Monika Kamińska for providing sky positions for satellites from STP. We are grateful to Dr. T.S. Kelso for development and maintaining of the CelesTrack.
This work is supported by National Key R\&D Program of China (Grant No. 2021YFA1600401/3), and by the Natural Science Foundation of China (Nos. 12090040/3, 12125303, 11733008).
Guoshoujing Telescope (the Large Sky Area Multi-Object Fiber Spectroscopic Telescope LAMOST) is a National Major Scientific Project built by the Chinese Academy of Sciences. Funding for the project has been provided by the National Development and Reform Commission. LAMOST is operated and managed by the National Astronomical Observatories, Chinese Academy of Sciences. The authors gratefully acknowledge the “PHOENIX Supercomputing Platform” jointly operated by the Binary Population Synthesis Group and the Stellar Astrophysics Group at Yunnan Observatories, Chinese Academy of Sciences. 
This research has made use of NASA’s Astrophysics Data System. It also made use of TOPCAT, an interactive graphical viewer and editor for tabular data \citep[][]{topcat}.

\section*{Data Availability}
The data underlying this article will be shared on reasonable request to the corresponding author.
LAMOST-MRS spectra are downloaded from \url{www.lamost.org}.




\bibliographystyle{mnras}



\appendix


\bsp	
\label{lastpage}
\end{document}